\documentstyle[epsfig,sprocl]{article}

\newcommand{\cpt}{{\chi_{PT}}}

\begin{document}

   
\title{Photonic Probes of the Deuteron in Effective Field Theory\footnote{Talk given at the Workshop on
    Chiral Dynamics 2000: Theory and Experiment, TJNL (Newport News, USA) 17th --
    22nd July 2000; to be published in the Proceedings. Preprint NT@UW-00-032.}}

\author{Silas R.~Beane}

\address{Department of Physics, University of Washington, \\
Seattle, WA 98195. }

\maketitle\abstracts{I give a taste of recent progress in using chiral
perturbation theory to understand nucleon and deuteron structure.}

%
\noindent A great deal of progress has been made unravelling nuclear
scales and constructing the effective field theory (EFT) relevant for
low-energy nuclear physics (for a recent review see
Ref.~\cite{reviews}). Here I will focus on a success and a puzzle.

\begin{figure}[!htb]
  \centerline{    \epsfig{file=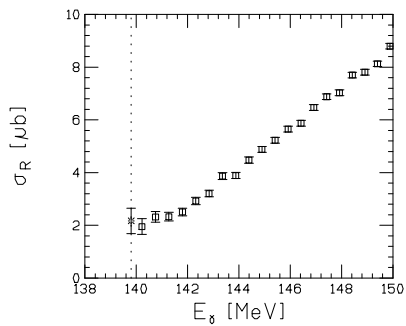,width=0.44\textwidth,clip=}}
\caption{
Total cross section for $\gamma d \to \pi^0 d$. The
data from SAL are depicted by the boxes, the CHPT threshold prediction (before
the data release) is the
star on the the dotted line (indicating the threshold photon energy).} 
\label{fig:pid}
\vspace*{-4ex}
\end{figure}

Neutral pion photoproduction on the deuteron has been computed to
$O(q^4)$ in chiral perturbation theory ($\cpt$)~\cite{silas2}. The {\it
prediction} for the threshold amplitude is $E_d = (-1.8 \pm 0.2)
\times 10^{-3}/m_{\pi^+}$.  To see the sensitivity to the elementary
neutron amplitude, we set it to zero and find $E_d = -2.6\times
10^{-3}/m_{\pi^+}$, which results in a cross-section twice as large as
the $\cpt$ prediction.  This is a unique situation as the EFT makes a
prediction that differs significantly from conventional models. A
measurement of the threshold amplitude has recently been carried out
at Saskatoon~\cite{jack}.  The experimental results for the pion
photoproduction cross-section near threshold are shown in
Fig.~\ref{fig:pid}, together with the $\cpt$ prediction at threshold.
At threshold, Ref.~\cite{jack} extracts $E_{d}=(-1.45\pm 0.09)\times
10^{-3}/m_{\pi^+}$.  While agreement with $\cpt$ to order $O(q^4)$ is
not better than a reasonable estimate of higher-order terms, it is
clearly superior to models.  This is compelling evidence of the
importance of chiral loops and of the consistency of nuclear EFT.

Electroproduction on the deuteron has recently been measured at
Mainz~\cite{aaron}. The two relevant multipoles have been computed to
$O(q^3)$ in the chiral expansion~\cite{ulf}. There are discrepancies
between data and theory. However, a true test awaits an $O(q^4)$
computation which is currently underway.

\begin{figure}[!htb]
  \centerline{\epsfig{file=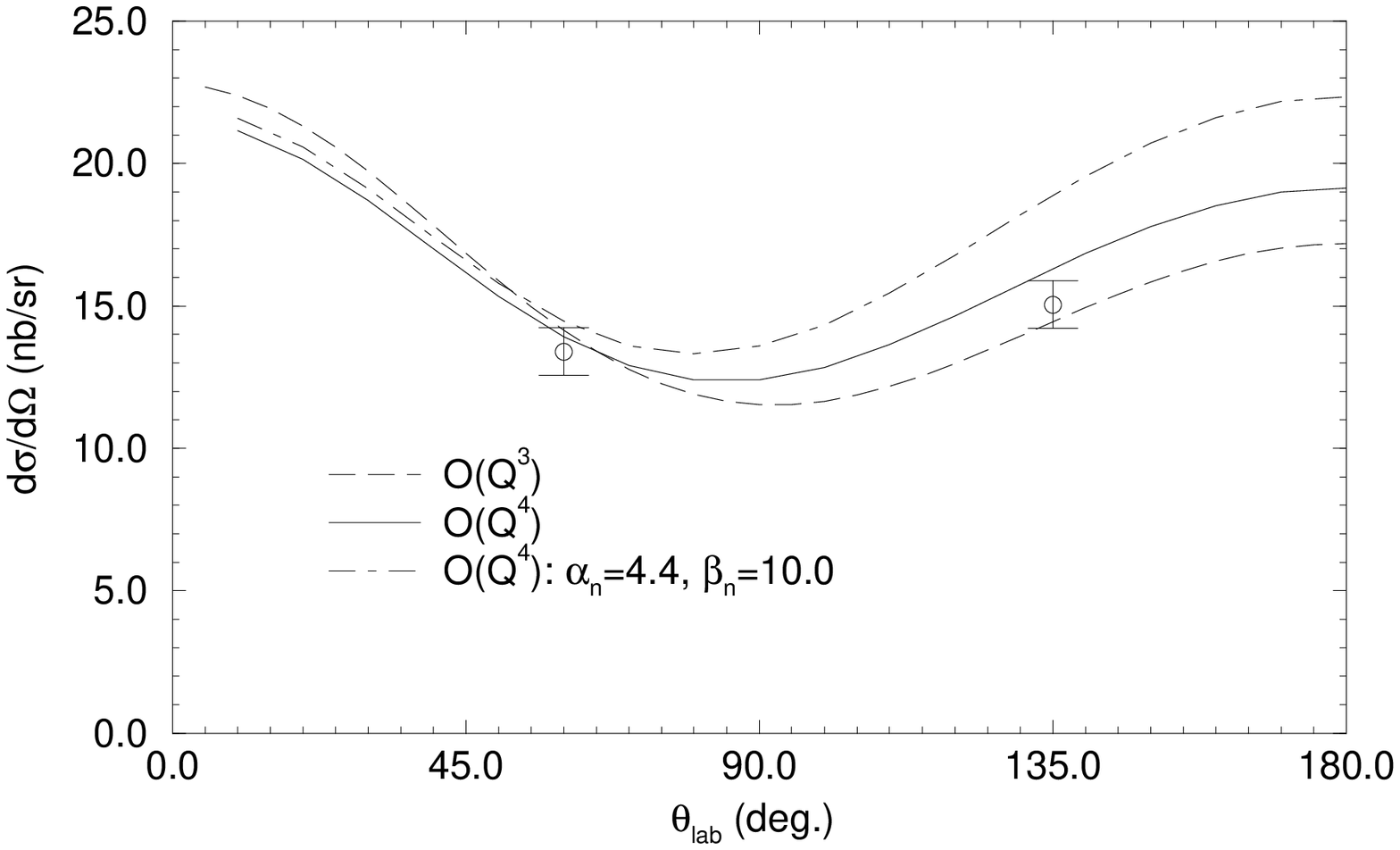,width=0.50\textwidth,clip=} \hfill
    \epsfig{file=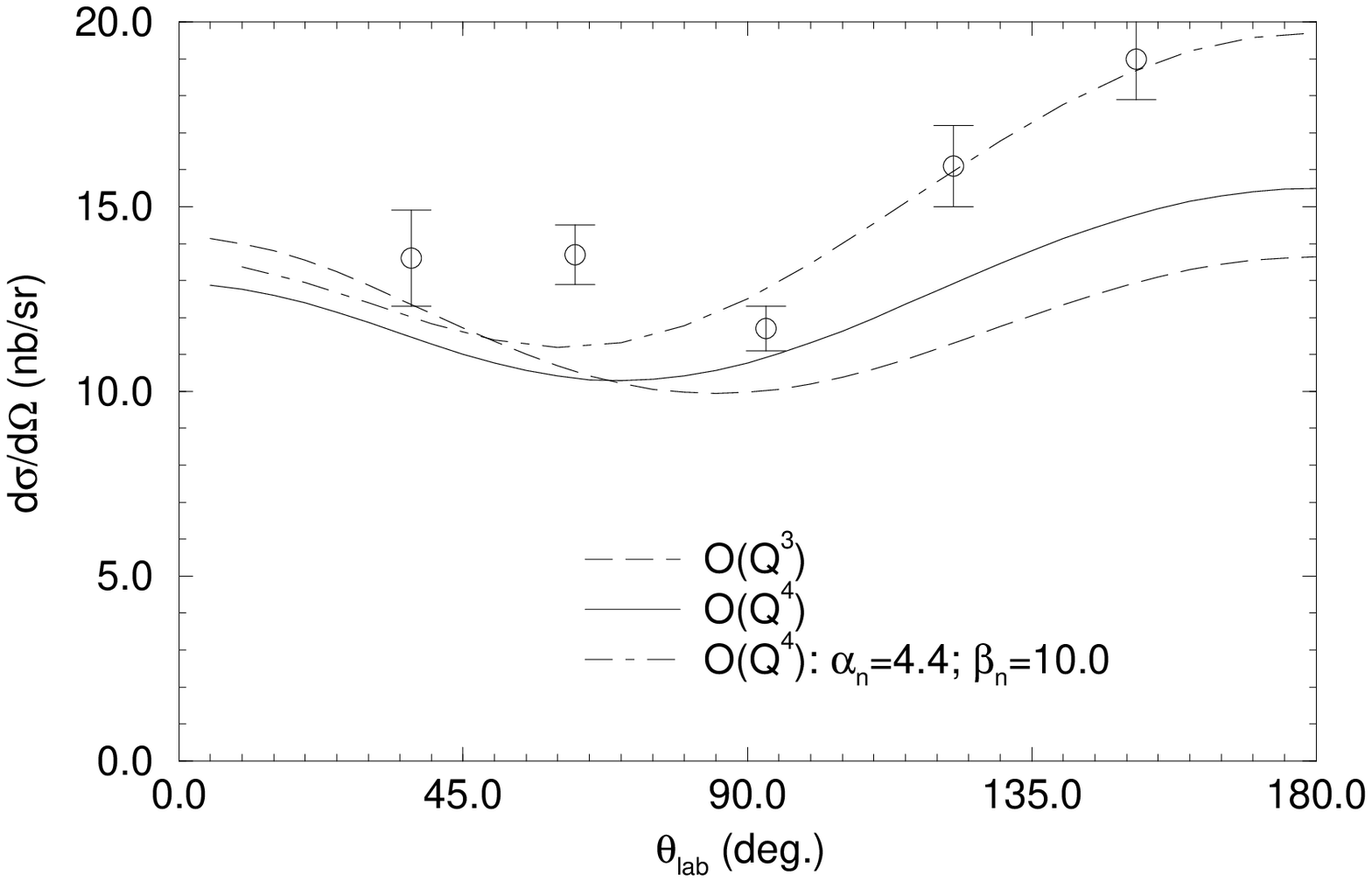,width=0.47\textwidth,clip=}}
\caption{The differential cross section for elastic 
  \protect$\gamma$-deuteron Compton scattering at incident photon energies of
  \protect$E_{\gamma}=69\ {\rm MeV}$ and \protect$95\ {\rm MeV}$ compared to
   data from Ref.~\protect\cite{lucas} and Ref.~\protect\cite{dave}, respectively.}
\label{fig:SAL}
\vspace*{-4ex}
\end{figure}

Compton scattering on the deuteron is important because it may provide
the only rigorous way to determine the polarizabilities of the
neutron. This process has been computed to $O(q^3)$ in
$\cpt$~\cite{silas1}. An $O(q^4)$ computation is currently underway.
The $O(q^3)$ result is in agreement with the $69~{\rm MeV}$ data (see
Fig.~\ref{fig:SAL}). There are $O(q^4)$ counterterms which shift the
nucleon polarizabilities. We use this freedom to fit the $95~{\rm
MeV}$ data using an incomplete $O(q^4)$ calculation.  A reasonable fit
at backward angles can be achieved with $\alpha_n =4.4$ and $\beta_n
=10$, which are in startling disagreement with the $O(q^3)$ $\cpt$
expectations. Also, the cross-section at 69 MeV with $\alpha_n =4.4$
and $\beta_n =10$ is then excessively forward peaked and misses the
data. This situation poses an interesting puzzle. Of course, a full
$O(q^4)$ calculation in $\cpt$ is necessary if the neutron
polarizability is to be extracted from the Saskatoon data in a
systematic way.

\end{document}